\documentclass{PoS}

\usepackage{setspace}

\def\hess{H.E.S.S.}

\def\gg{$\gamma$}
\def\gr{$\gamma$-ray}
\def\grs{$\gamma$~rays}

\def\deg{$^{\circ}$}

\def\diffi{\emph{cm$\textsuperscript{-2}$.s$\textsuperscript{-1}$.TeV$\, \textsuperscript{-1}$}}

\def\xeff{$X_{eff}$}
\def\pdf{\emph{PDF}}
\def\pdfs{\emph{PDFs}}
\def\MSSG{\emph{MSSG}}

\def\phi12{$\Phi$(1TeV).\emph{10$\textsuperscript{-12}$}}

\def\pks{~\mbox{PKS 2155-304}}

\title{Xeff analysis method optimization to enhance IACTs performances}

\ShortTitle{XEff analysis method optimization}

\author{\speaker{C.~Trichard}\\
LAPP, Université Savoie Mont Blanc, CNRS/IN2P3, Annecy-le-Vieux, France\\
        E-mail: \email{trichard@lapp.in2p3.fr}}
\author{A.~Fiasson\\
LAPP, Université Savoie Mont Blanc, CNRS/IN2P3, Annecy-le-Vieux, France\\
        E-mail: \email{fiasson@lapp.in2p3.fr}}
\author{G.~Maurin\\
LAPP, Université Savoie Mont Blanc, CNRS/IN2P3, Annecy-le-Vieux, France\\
        E-mail: \email{maurin@lapp.in2p3.fr}}
\author{G.~Lamanna\\
LAPP, Université Savoie Mont Blanc, CNRS/IN2P3, Annecy-le-Vieux, France\\
        E-mail: \email{lamanna@lapp.in2p3.fr}} 

\abstract{
The seek of high precision analyses in \gr\ astronomy leads to the implementation of multivariate combination, benefiting from several reconstruction methods. Such analysis, called \xeff\ , was developed for the \hess\ data using three shower reconstruction methods. This paper presents the improvement granted to this analysis by refining the distribution calculation of discriminant variables, considering observation conditions, and adding new variables in the \xeff\ combination. The efficiency of the analysis is presented using simulations and real data. A comparison with the standard analysis {\it model++}, for a typical set of sources, shows a significant gain in sensitivity.   
}

\FullConference{The 34th International Cosmic Ray Conference,\\
		30 July- 6 August, 2015\\
		The Hague, The Netherlands}

\begin{document}

\section{Introduction}

Identification of photons against hadron cosmic rays is very challenging for Imaging Air Shower Telescopes such as \hess . A multivariate method, \xeff , has been developed to improve this separation~\cite{Dubois09}. The aim is to collect the maximum information provided by the air shower reconstruction and the Monte-Carlo simulations for photons and hadrons. Originally the \xeff\ analysis combined three reconstruction methods, the reconstruction of Hillas parameters, and two 3D reconstructions: a semi-analytical one, referred hereafter as {\it model} and its improved version {\it model++}~\cite{DeNaurois09}, and an analytical one, referred as {\it 3D-model}~\cite{Lemoine06}. Since the {\it model} reconstruction provides a very good angular resolution and energy reconstruction, it has been used and \xeff\ serves as an estimator to separate \gg\ and hadron events. The performance resulting from the combination of {\it Hillas} parameters, {\it 3D-model}, the latest version of {\it model}  ({\it model ++}) and other discriminating variables are reported below.

\section{The \hess\ data analysis methods}
The three shower reconstruction methods applied so far in the HESS data analysis are briefly described in this section.

\subsection{Hillas analysis}

The first method applied to the \hess\ data makes use of the {\it Hillas} parameters~\cite{Aharonian06a}. These parameters are extracted by fitting an ellipse to the image. The direction of the particle is determined by the orientation of the ellipse and its energy by both the total image amplitude and the reconstructed impact parameter of the shower.
The hadron/\gg\ discrimination is performed using scaled variables. The length and width (geometric parameters of the fitted ellipse) are scaled with the mean value given by Monte-Carlo simulations.  

\subsection{3D model analysis}

A 3D-modeling~\cite{Lemoine06} of the air showers has been developed to benefit from the array of telescopes through the stereoscopic information. This parametrization leads to a quantity called "3D-width". This parameter of the fit refers to the transverse standard deviation of the Gaussian distribution of the shower with respect to the main axis. This variable has a discrimination power between \grs\ and hadron induced showers. Scaling the 3D-width by the depth of shower maximum ({\it reduced 3Dwidth}), enhances the background rejection capability. 

\subsection{Semi-analytical model analysis}

The semi-analytical {\it model} analysis has been developed for \hess\ data analysis~\cite{DeNaurois09}. This method compares the images of the atmospheric showers to predicted images given by a semi analytical {\it model}. This prediction is stored in a look-up table and a log-likelihood maximization is done over all the pixels. The parameters of the primary particle are given by the most probable set of images. The background and signal separation is performed through the goodness-of-fit variable. Then this variable is rescaled using the same method as for the {\it Hillas} analysis.   

\section{The \xeff\ analysis}\label{secxeff}

\subsection{\xeff\ estimator}

The discriminating parameters, described above, are already powerful by themselves and applied to the \hess\ data give already good background rejection for bright sources. The aim of this approach is to combine the information produced by the three analysis methods. All image information will be collected in a single discriminating estimator, \xeff. This variable, associated to each event, has the power of an event-by-event gamma-mistag probability estimator. The definition of this estimator, introduced by~\cite{Buskulic96}-~\cite{Jaffe}, follows the relation $$X_{eff}=\frac{\eta \prod_j{h_j(x_j)}} {\eta \prod_j{h_j(x_j)} + (1-\eta)\prod_j{g_j(x_j)}} $$
where $h(x_j)$ and $g(x_j)$ are the probability density functions (\pdf) of variable $j$ for hadron and \gg\ events respectively, $\eta=\frac{N_b} {N_b+N_{\gamma}}$ is the assumed relative background fraction events, $N_b$ and $N_{\gamma}$ being the number of background and \gg\ events in data sample.

\subsection{Input variables for combination}

The first version of the \xeff\ analysis within the \hess\ analysis framework~\cite{Dubois09} used four variables : the mean scaled width and length from {\it Hillas} reconstruction, the rescaled width from {\it 3D-model} and the mean scaled goodness from {\it model} reconstruction. In this updated version of \xeff\ the latest variable from {\it model}, the mean scaled shower goodness (\MSSG\ see~\cite{DeNaurois09}), has been substituted to the mean scaled goodness. 

There are almost no correlation for \grs\ and partial correlation for hadrons between the mean scaled shower goodness and the other variables described above. Thus the \MSSG\ can be used efficiently as a variable in \xeff\ analysis. Similar results have been obtained by~\cite{Dubois09} between the remaining variables used in this analysis. Adding \MSSG\ instead of Goodness increases significantly the discrimination power. In order to quantify the benefits of taking into account the reconstructions other than {\it model++} in this multivariate analysis, the performance of \xeff\ will be compared in the next sections to the {\it model++} analysis.

In order to improve the discriminant power of the analysis, 3 variables were added. The directions reconstructed are slightly different for one method to another. These differences spread to higher values for hadrons~\cite{Dubois09} providing a discrimination power. Therefore, the differences between reconstructed directions ($\Delta\theta_{\mathrm{Hillas-Model}}$, $\Delta\theta_{\mathrm{Hillas-Model3D}}$ and $\Delta\theta_{\mathrm{Model-Model3D}}$) have been introduced as
supplementary PDFs in the \xeff\ estimator.

\subsection{Training strategy}
\label{sec-trainstrat}

The approach of multivariate analysis is to use all information provided by the various shower reconstructions. To reduce the simulation uncertainties, we used real data to determine the hadron distributions for each variable. A sample of data (obtained in 2004 and 2005) without \gr\ sources in the field of view ("\emph{OFF}") and far from the galactic plane, to avoid the galactic \gr\ background, were taken. These data come from AGN observations at different zenith angles and different distances from the center of the telescope cameras (offset). 

Unfortunately there are no free background \gr\ sources available in the \hess\ data. However it has been shown that \gr\ simulations reproduce faithfully the data. We used Monte-Carlo simulations with an offset from 0.5\deg\ to 2.5\deg\ to compute \gr\ distributions.

The \pdfs\ computation is different from the previous version of \xeff\ analysis. In order to compute the \pdf\ as close as possible to the real data, our simulation and \emph{OFF} data samples have been separated in 6 energy bins and 7 zenith angle bins. Air shower development slightly depends on the azimuth angle. The simulation sample has been further separated into two different sets corresponding to azimuth angle of 0\deg\ and 180\deg. The lack of statistic prevents making such separation for hadrons.  Small differences were expected for the \emph{mean scaled} variables but the $\Delta\theta$ distributions should strongly depend on the energy and zenith angle. The aim is to select the \pdfs\ corresponding to the observation conditions to perform the analysis. Including these customizations in the training improves the discrimination power of \xeff.
This training method increases significantly the number of \pdfs\ to compute, up to 882, which requires an automated procedure for a proper derivation from the original MC or OFF data distribution. The kernel estimation method from the package \emph{RooFit} has been used. It ensures a determination of PDFs close to the original distribution without assumption on their shapes.

\subsection{Event preselection}
A set of pre-cuts is usually applied in the \hess\ data analysis for two reasons : to select events that will be correctly reconstructed and to reject a significant part of hadron events without rejecting \gg\ events. The analysis presented here uses the {\it model} reconstruction. The same pre-selection cuts as applied for the {\it model} analysis were chosen. They comprise the distance from the image barycenter to the camera center, the minimal charge of the image and the likelihood in the pure NSB hypothesis.
It has been shown that the primary interaction of the particle scaled in terms of photon radiation length could be used as a pre-cut to reject an important fraction of hadron events with a low impact on \grs\ ~\cite{Dubois09}. Events with a reconstructed primary depth between -1 and 4 were kept. Events with values out of selection range for at least one of the variables were flagged as background.

\subsection{Simulated efficiencies}

\begin{figure}[!ht]
 \centering
 \includegraphics[width=0.40\textwidth]{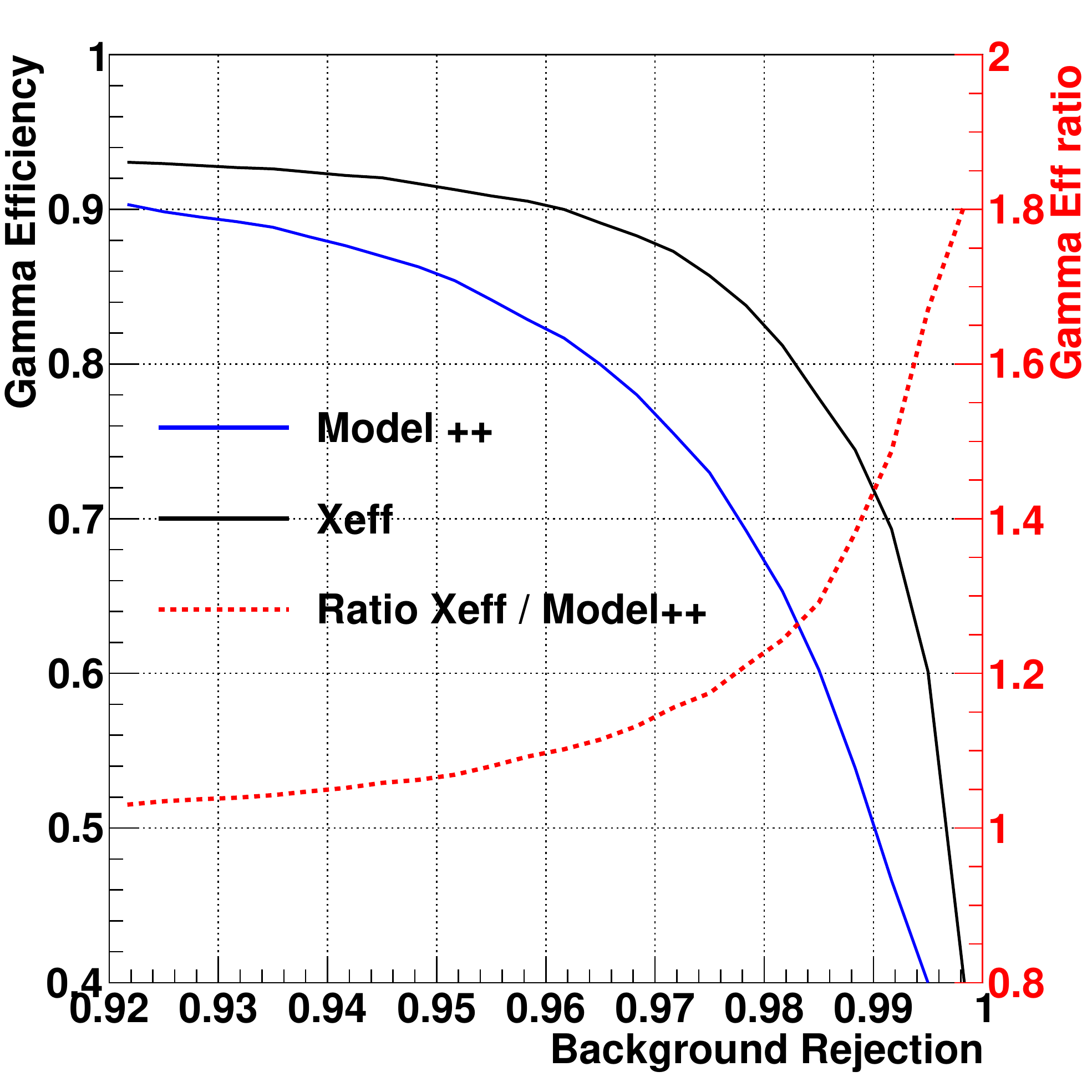} 
   \caption{The blue and black lines show the \gg\ efficiency of {\it model}~\cite{DeNaurois09} and \xeff\ respectively for a zenith angle between 15\deg\ - 25\deg, an energy between 2 - 5 TeV and a muon efficiency of 100\%. The ratio of \xeff\ \gg\ efficiency over {\it model}'s is shown in red. The scale is at the right side of the plot.  
  \label{GamvsBack}}
\end{figure}

To probe the performance of \xeff\ applied on \hess\ data we analyzed simulated \gg\ events and OFF data. Figure~\ref{GamvsBack} shows an example of \gr\ efficiency versus the background rejection for several Zenith and Energy bins. A comparison with the {\it model++} analysis, which is generally used for \hess\ data, is added. \xeff\ has always greater \gg\ efficiency than {\it model} for a given background rejection in each bin. The standard \hess\ analyses reject around 95\% of background. For such rejection, \xeff\ provides $\sim$10$\%$ more \gg\ events than {\it model++} (see Figure~\ref{GamvsBack}). 

\subsection{Determination of discrimination cuts}
Once an event passed through the pre-selection described above, we still have to determine the nature of the primary particle. We use the \xeff\ value to flag this event as hadron/\gr-like. The cut to apply on the \xeff\ estimator is arbitrary.

According to the definition of \xeff, $\eta$ is a purity estimator. It extends from 0 for background-free to 1 for no \gg\ events data samples respectively. Four values for $\eta$ have been defined depending on the brightness of the source. This provides 4 different analyses adapted for specific source type. 
A cut on the \xeff\ variable is then applied. A choice has been done to keep at least the same \gg\ efficiency as {\it model++} across energy and zenith ranges as wide as possible. According to simulations, a single cut value of 0.3 is applied for all values of $\eta$ (i.e. source brightness). This provides a good compromise between significance of the signal and \gr\ excess. Applying a unique cut for all $\eta$ values leads to different background rejection and \gg\ efficiency. Indeed, high value of $\eta$ (e.g. 0.7) corresponds to high background rejection whereas for $\eta=0.1$ the \gg\ efficiency is at maximum. So for faint sources, the \xeff\ analysis focuses on signal significance enhancing the detection capability, while for bright sources the \gr\ excess is favored. The \xeff\ analysis parameters are summarized in Table~\ref{tabcuts}.

\begin{table}[!ht]
 	 \caption{Summary of cuts applied in the \xeff\ analysis depending on the brightness of the source.}
	\begin{center}
  	\begin{tabular}{|c|ccc|}
	
  	\hline
	Brightness &\phi12 & $\eta$ & \xeff\ \\
	&\diffi&& \\
	
	\hline
	\hline
	
	Very Bright & $>10$ & 0.1 & $<0.3$  \\
	
	Bright &  1 ... 10  & 0.2 & $<0.3$  \\

	Medium &  0.5 ... 1  & 0.4 & $<0.3$\\

	Faint &  $< 0.5 $  & 0.7 & $<0.3$  \\

	\hline

   	\end{tabular}
 		\end{center}
		\label{tabcuts}
\end{table}

\subsection{\gg\ - hadron efficiencies}

We estimated the \gg\ efficiencies using Monte-Carlo \grs\ for the \xeff\ cuts defined in previous section. We used the bins defined above for the computation of the \pdfs . The \gg\ efficiency for $\eta=0.1$ is $>$85\% for most part of Zenith/Energy bins. The \gg\ efficiency for standard \hess\ analyses is generally $<$70\%. While for $\eta=0.7$ the lower \gg\ efficiency ($>$60\% in most bins) is comparable to other \hess\ analyses.
For this configuration the high level of hadron rejection ($>$96\%) and a good \gg\ efficiency provide a powerful analysis to extract faint signals and claim new detections.     

\subsection{Effective Area}

\begin{figure}[!ht]
 \centering
\includegraphics[width=0.41\linewidth]{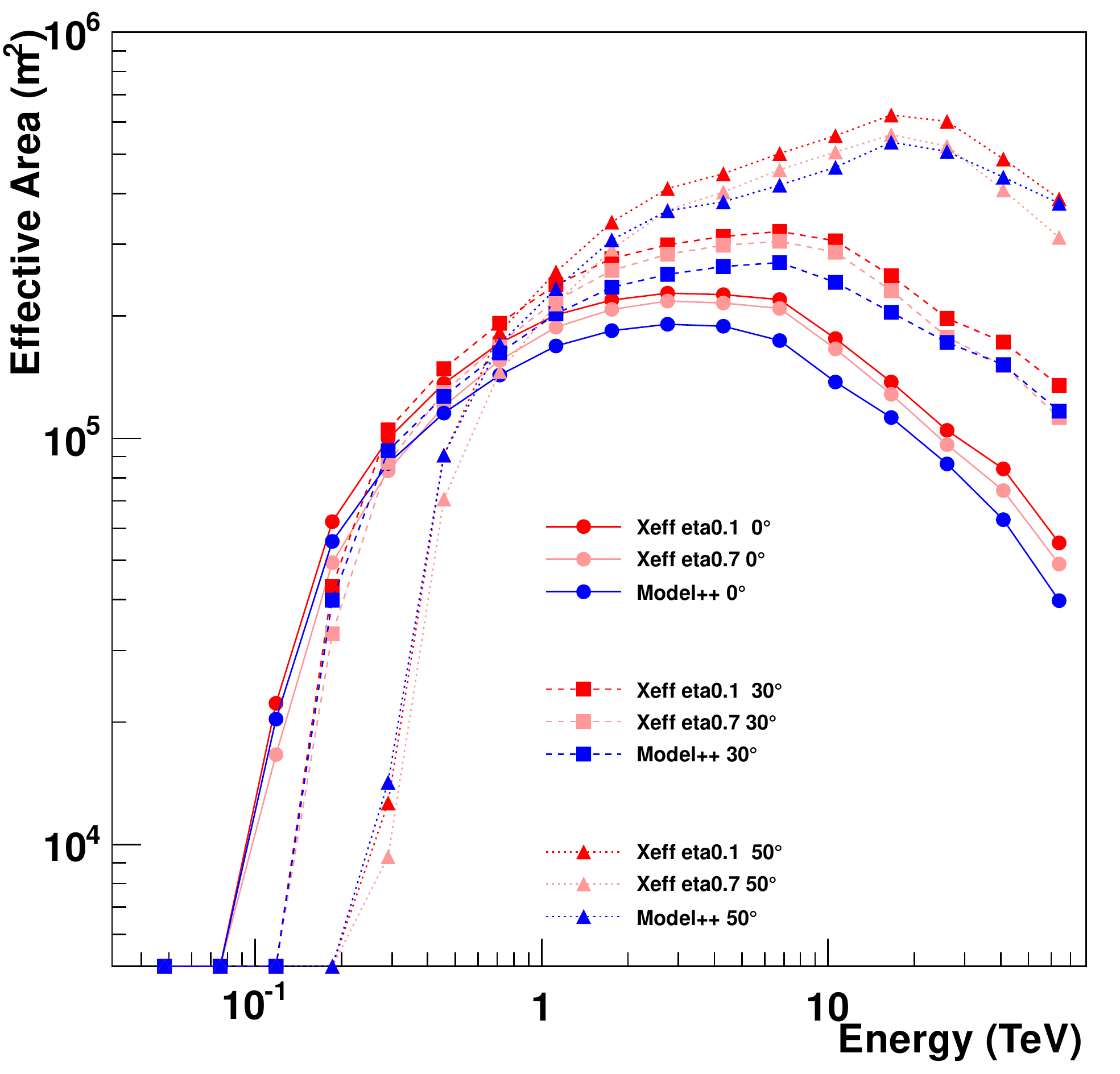} 
 \includegraphics[width=0.58\linewidth]{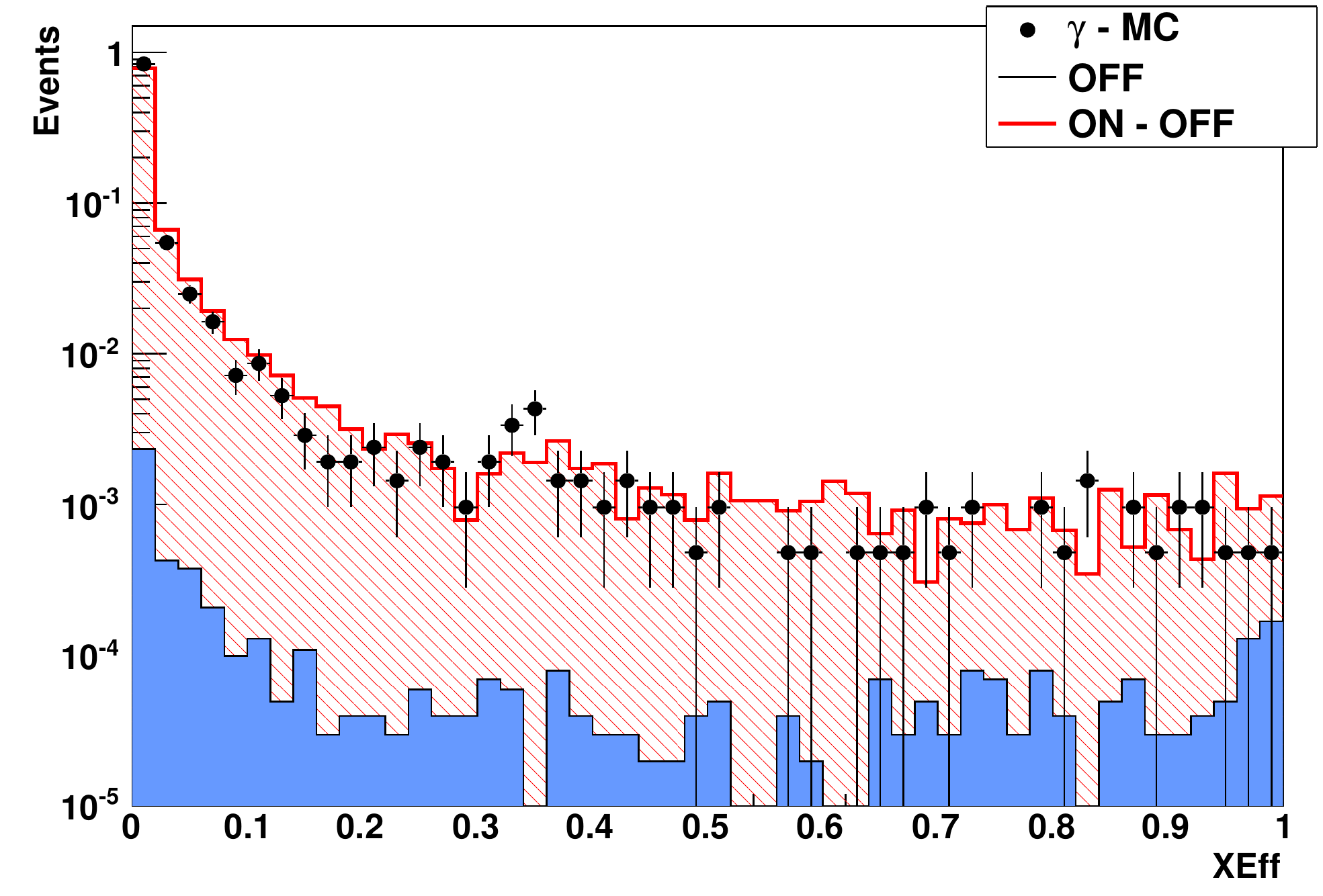} 
  \caption{ Left: Effective area of \xeff\ analysis for $\eta=0.1$ and $0.7$, and for 3 zenith angles.  The effective area of {\it model++} analysis is added for comparison~\cite{DeNaurois09}. Right: \xeff\ distribution obtained using 3 runs of \pks's flare and Monte Carlo simulations~\cite{PKSpub}. The red line represents the ON events normalized after background subtraction. The black points show the normalized distribution obtained with Monte-Carlo simulations corresponding to the same observation conditions. The blue histogram is the normalized background distribution.\label{MCdata}
  \label{Accep}}
\end{figure}

Figure~\ref{Accep} presents the effective area of \xeff\ and {\it model++} analyses. The results for $\eta=0.1$ and $0.7$ and for 3 zenith angles are shown. \xeff\ analysis provides a better \gg\ acceptance for almost all the \hess\ energy range and for all zenith angles. As expected the \xeff\ effective area is lower for $\eta=0.7$ but is still better or similar to {\it model++}'s. The splitting of data and MC samples during the training of \xeff, see section~\ref{sec-trainstrat}, does not introduce visible effects on the effective area shape. Theses results demonstrate the gain in sensitivity provided by the multivariate \xeff\ analysis compared to the standard most powerful \hess\ analysis.

\section{Systematic studies}

\subsection{Comparison between simulation and real data}

\hess\ data are contaminated by hadron induced air showers. Furthermore, no background-free observations can be done. To control the behavior of the analysis on \gr\ data, we can use the July 2006 \hess\ observation of \pks's flare~\cite{PKSpub}. Thanks to the exceptional \gr\ flux, this flare produced the purest data set available which can be directly used for a comparison with Monte-Carlo simulations.

Three observations of July 28$^{th}$ 2006 corresponding to \pks's flare were used. Figure~\ref{MCdata} right shows the \xeff\ normalized distributions for this analysis. The very low level of background events provides a very high purity of the sample. These data were compared to simulations corresponding to the same zenith and target offset. However, given the available MC statistics, simulations for azimuth angles of 0\deg\ and 180\deg\ and spectral index from 3.2 to 3.6 (a spectral index of $\sim$3.4 was measured on the source with other analysis~\cite{PKSpub}) have been summed. The comparison shows that Monte-Carlo simulations are well describing the observed \grs\ and the small differences can be understood as statistical fluctuations and possible bias due to Monte Carlo selection.

\subsection{Comparison with {\it model} and published results on test sources}

Simulations have shown a significant gain of \xeff\ analysis compared to standard analyses. Results obtained with \xeff\ and {\it model} analyses on several well known and/or published sources were compared. The sources have been arbitrarily chosen to be as representative of \hess\ source type as possible : bright and faint, galactic and extragalactic, and soft and hard spectral index sources. As expected for faint sources, \xeff\ provides a gain of significance of $\sim$7$\%$ up to $\sim$25$\% $ while for bright sources the gain is significant for the \gr\ excess $\sim$10$\%$, up to $\sim$17$\%$. The differences from one source to another is due to the different observation conditions (in agreement with simulations).

\subsection{Spectral comparison}

The cuts introduced by \xeff\ analysis provided a better background rejection and/or a better \gg\ efficiency than standard \hess\ analyses. The spectral results may be affected by the event selection. A spectral analysis of the same set of sources as previously used was performed, assuming that the spectra are following a power law. Consistent results were obtained and no bias was introduced by the \xeff\ method.  

The fits with exponential cut off power law model (for sources with such spectral shape) give consistent results as well. Finally, a comparison with published results using other analyses ({\it Hillas}, {\it 3D-model} ...) shows consistent spectral parameters.  
  
\xeff\ analysis depends on the free parameter $\eta$ . Its value affects significantly the background rejection and the \gg\ efficiency. The influence of the choice of this parameter on the results and in particular the spectral behavior of the analysis has been studied. Table~\ref{cutstudies} reports the results for 3 sources with different brightness (assuming the spectra follow a pure power-law). No evidence of significant bias or correlation between the value of $\eta$\ and the spectral parameters has been observed. This important feature leads to the conclusion that the parameter of \xeff\ analysis can be selected according to the purpose of the analysis. One can select hard cuts (high value for $\eta$) to get a good signal significance or loose cuts (low value for $\eta$) to increase \gr\ statistics.

\begin{table}[!h]
\tiny
	\caption{Results of \xeff\ analysis for 3 sources with different level of brightness. A pure power-law has been fitted on each spectrum. Various values of the free parameter $\eta$ have been tested for each source. Results published with independent analyses are provided when available \cite{ls5039}\cite{g09}\cite{cenA}. Only statistical errors are indicated.}
	\label{cutstudies}

	\begin{center}
  	\begin{tabular}{|cccccccc|}
  	\hline
	Source & Method & $\eta$ & N$_{\gamma}$  & N$_{\sigma}$ & $\Gamma$ &\phi12& \\
	&&&&&&\diffi&\\
  	\hline
	\hline

LS5039 & Xeff & 0.1 &  3445 & 57.1 & $ 2.27 \pm 0.03 $ & $ 1.74 \pm 0.04 $ &\\  
&Xeff & 0.2  & 3360 & 57.8 & $2.27 \pm 0.03$ & $1.72 \pm 0.04$ &\\ 
&Xeff & 0.4 &  3237 & 58.9 & $ 2.27 \pm 0.03 $ & $1.70  \pm 0.04 $ &\\  
&Xeff & 0.7 &  2991 & 59.8 & $ 2.25 \pm 0.03 $ & $1.69  \pm 0.04 $ &\\  
&&&&&&&\\
G0.9+0.1 & Xeff & 0.1 &  607 & 18.4 & $2.34 \pm 0.07 $ & $ 0.79 \pm 0.05 $ &\\  
&Xeff & 0.2 &  563 & 17.9 & $2.30 \pm 0.07 $ & $ 0.77 \pm 0.05 $ &\\  
&Xeff & 0.4 & 533 & 18.2 & $2.30 \pm 0.07$ & $0.78 \pm 0.05$& \\ 
&Xeff & 0.7 &469 & 18.3 & $2.26 \pm 0.07 $ & $0.75  \pm 0.05 $ &\\  
&Pub   & - &  - &-  & $2.4 \pm 0.1 $ & -& \\

&&&&&&&\\
Cen A & Xeff& 0.1 & 410 & 8.6 & $ 2.6 \pm 0.2 $ & $0.26  \pm 0.03 $& \\  
&Xeff& 0.2 &  427 & 9.4 & $ 2.7\pm 0.2 $ & $ 0.26 \pm 0.03 $& \\  
&Xeff& 0.4 &  414 & 9.9 & $ 2.7 \pm 0.2 $ & $0.26  \pm 0.03 $& \\  
&Xeff & 0.7 & 385 & 10.9 & $2.5 \pm 0.2$ & $0.28 \pm 0.03$ &\\ 
&Pub   &- &- &-  & $2.7 \pm 0.5$  & $0.25 \pm 0.05$& \\
\hline

 	\end{tabular}
  		\end{center}
\end{table}

\section{Summary}

The \xeff\ multivariate analysis, already used for \hess\ data analysis, has been improved by taking into account the mean zenith and azimuth angles of the run and the energy of the event. Incorporating the differences between the reconstructed directions of {\it model}, {\it Hillas} and {\it 3D-model} reconstructions increases the discrimination power of the \xeff\ combination. Simulation studies show that a higher \gg\ efficiency and a better background rejection than standard \hess\ analyses can be reached. The free parameter $\eta$ allows to set different background rejection levels depending on the brightness of the target. This feature is a powerful way to increases the significance of faint sources enhancing the detection probability. Looking at various well known sources, it has been shown that \xeff\ produces a higher \gr\ excess for bright sources of $\sim$10\% and a higher significance of $\sim$7\% for faint sources, compared to the powerful {\it model} analysis. The present work provides a powerful analysis tool, suitable for all kind of physics analysis feasible with \gr\ imaging Cherenkov telescopes. 
\\
\\

{\bf Acknowledgements:} The authors would like to thank the H.E.S.S. Collaboration for the technical support and fruitful discussions, as well as for allowing us to use H.E.S.S. data in this publication.

\end{document}